\documentclass[aps,prl,twocolumn,nofootinbib,longbibliography,superscriptaddress]{revtex4-2}

\usepackage[dvipdfmx]{graphicx}
\usepackage{amsfonts}
\usepackage{amscd}
\usepackage{amsmath}
\usepackage{amssymb}
\usepackage{newtxtext,newtxmath}
\usepackage{here}
\usepackage{tabularx}
\usepackage{etoolbox}
\usepackage{bm}
\usepackage{mathrsfs}
\usepackage{listings}
\usepackage{color}
\usepackage{braket}
\usepackage[normalem]{ulem}  
\usepackage{hyperref}

\newcommand{\nn}{\nonumber}
\newcommand{\beq}{\begin{eqnarray}}
\newcommand{\eeq}{\end{eqnarray}}

\newcommand\erase{\bgroup\markoverwith{\textcolor{blue}{\rule[.5ex]{2pt}{1pt}}}\ULon}

\begin{document}
\title{Microwave Kerr/Faraday Resonance in Two-dimensional Chiral Superconductors}

\author{Taiki Matsushita}
\affiliation{Department of Materials Engineering Science, The University of Osaka, Toyonaka, Osaka 560-8531, Japan}
\affiliation{Spintronics Research Network Division, Institute for Open and Transdisciplinary Research
Initiatives, The University of Osaka, Japan}
\author{Jun'ichi Ieda}
\affiliation{Advanced Science Research Center, Japan Atomic Energy Agency, Tokai, Ibaraki 319-1195, Japan}
\author{Yasufumi Araki}
\affiliation{Advanced Science Research Center, Japan Atomic Energy Agency, Tokai, Ibaraki 319-1195, Japan}
\author{Takahiro Morimoto}
\affiliation{Department of Applied Physics, The University of Tokyo, Hongo, Tokyo,
  113-8656, Japan}
\author{Ilya Vekhter}
\affiliation{Department of Physics and Astronomy, Louisiana State University, Baton Rouge, LA 70803-4001, USA}
\author{Youichi Yanase}
\affiliation{Department of Physics, Graduate School of Science, Kyoto University, Kyoto 606-8502, Japan}
\date{\today}

\begin{abstract}
We investigate the polar Kerr and Faraday effects in two-dimensional multiband chiral superconductors.
We show that the clapping modes--the relative phase and amplitude oscillations between two chiral components of the superconducting order parameter--lie well within the quasiparticle excitation gap in multiband systems and dominate these magneto-optical responses in the microwave regime.
The Kerr and Faraday rotation angles exhibit the resonant enhancement with sign reversals in the microwave regime as a function of the light frequency, reaching peak values on the order of 100 nrad--10 $\mu$rad in thin films of candidate chiral superconductors.
These resonances are accessible in superconducting atomic layer materials and provide a generic probe of chiral superconductivity in two-dimensional systems.
\end{abstract}
\maketitle

{\it Introduction.---}
Detecting time-reversal symmetry breaking (TRSB) is one of the central issues in superconductivity research~\cite{ghosh2020recent}. 
Among TRSB superconductors, a chiral order parameter, $\Delta(\bm k) \propto (k_x \pm i k_y)^{m_z}$ ($m_z \in \mathbb{Z}$, $m_z \neq 0$), describes a condensate with a polarized orbital angular momentum and is a prototypical such topological superconducting phase~\cite{xiao_TSC,tanaka2011symmetry,kallin2016chiral,sato2016majorana,mizushima2016symmetry,sato2017topological}.
In topological superconductors, Majorana zero modes emerge as zero-energy Bogoliubov quasiparticles at vortices and boundaries, and their non-Abelian statistics offers a route to decoherence-free qubits~\cite{alicea2012new,sannnomajorana,machida2019zero}.
Accordingly, identifying topological superconductivity -- or its parent chiral superconducting order -- is an urgent priority.

Although several candidate materials have been proposed, definitive evidence of chiral superconductivity remains elusive.
For example, the (even-parity) $d$-wave chiral order has been proposed in Sr$_2$RuO$_4$, URu$_2$Si$_2$, SrPtAs, and some kagome metals~\cite{xia_PKE,maeno2024still,shemm_PKE_uru2si2,yamashita2015colossal,sumiyoshi2013quantum,kasahara2009superconducting,kittaka2016evidence,Biswas_SrPtAs,Fischer_SrPtAs,tazai2025chiralnematic,yoshida2025_ATHE,Le2024}.
UPt$_3$ and U$_{1-x}$Th$_x$Be$_{13}$ are candidates odd-parity chiral superconductors~\cite{upt3review,adenwalla1990phase,nomoto_upt3,hasselbach1989critical,yanaseupt3,schemm2014observation,izawa2014pairing,tsutsumi2013upt3,upt3_josephson,rauchschwalbe1987phase,heffnerube13,shimizu2017quasiparticle,sigristube13,mizushima2018topology}.
More recently, chiral superconductivity has been suggested in atomic layer materials, such as homobilayer/twisted-bilayer transition metal dichalcogenides (TMDC) and rhombohedral graphene~\cite{kanasugitmd,yuantmd,wutmd,geier2024chiral, Han2025}.

In chiral states, the polarized orbital angular momentum of Cooper pairs breaks the mirror reflection symmetry together with time-reversal symmetry, thereby allowing zero-field transverse (Hall) responses.
However, in superconductors, the supercurrent shunts a zero-frequency (dc) electrical Hall response, and thus these broken symmetries can be most clearly detected via the anomalous thermal Hall effect (ATHE)~\cite{readFQHE,nomuracross,sumiyoshi2013quantum,moriya2022intrinsic,yip2016low,yilmazATHE,waveATHE,ngampruetikorn2024anomalous,matsushita2024impurity}.
In principle, the ATHE can serve as a direct probe for the TRSB topological phase because the topological contribution to the zero-field thermal Hall response is quantized at low temperatures~\cite{readFQHE,sumiyoshi2013quantum,sato2016majorana}, although the impurity-induced (non-quantized) contributions are also substantial~\cite{yip2016low,yilmazATHE,waveATHE,ngampruetikorn2024anomalous,matsushita2024impurity}.
Nevertheless, its observation remains extremely challenging due to the weak thermal Hall signal, while the ATHE has recently been observed in a kagome superconductor~\cite{yoshida2025_ATHE}.

Broken time-reversal and mirror-reflection symmetries can also be detected by the polar Kerr effect (PKE), which vanishes whenever the zero-field Hall response is absent [Eq.~\eqref{kerrangle}]~\cite{goryo2008impurity,wangkerr,furusaki_AHE,liu_extrinsicAHE, Lutchyn_extrinsicAHE,edward_intrinsicAHE,konigkerr,wangkerr,xia_PKE,li_extrinsicAHE, Yazdani_intrinsicAHE, Yakovenko_AHE, maeno2024still,shemm_PKE_uru2si2,schemm2014observation,hayes_kerr,ajeesh_kerr}.
Moreover, the PKE is accessible in atomic layer materials~\cite{huang2017layer,shimano2013quantum,fei2018two,zhu_kerr,lee2017valley,okada_kerr}.
Its application to superconducting atomic layer systems is especially significant, given the present lack of reliable probes of spontaneous TRSB in such systems.

Collective excitations reflect the symmetry structure of complex ordered states.
The clapping modes--the relative amplitude and phase oscillations between the two chiral components (e.g., $k_x\pm ik_y$)--serve as a fingerprint of chiral superconductivity~\cite{vollhardt2013superfluid}.
In single-band models, the clapping modes lie well below the quasiparticle continuum and are long-lived~\cite{sauls_clapping,matsushitaAAEE}, whereas in multiband systems, both their stability (lifetime) and magneto-optical responses (such as PKE) remain unexplored~\cite{niederhoff2025current}.
For many candidate chiral superconductors, their energies lie in the microwave regime.
So far, the PKE has been observed at near-infrared frequencies~\cite{xia_PKE,shemm_PKE_uru2si2,schemm2014observation,hayes_kerr}.
In those experiments, the photon energies far exceed the superconducting gap and thus cannot resolve the collective excitations.
Recent advances make magneto-optical measurements feasible in the microwave range, raising the question of the influence of collective modes on the PKE~\cite{roppongi2024microwave,chouinard_microwave,chouinard2025microwave}.

In this letter, we demonstrate the stability (long lifetimes) of clapping modes and propose the collective mode mechanisms for the PKE and the Faraday effect (FE) in multiband chiral superconductors.
This finding sharply contrasts with previous analysis within the single-band model, in which the clapping modes do not contribute to these magneto-optical responses~\cite{yip1992circular}.
We show that in multiband chiral superconductors, the clapping modes are responsible for the magneto-optical responses in the microwave regime, with resonant enhancement when the light frequency coincides with the energies of the clapping modes.
The resulting Kerr and Faraday rotation angles reach the order of 100 nrad--10 $\mu$rad in thin-films, well within the experimental sensitivity.
We further emphasize that these signals are accessible in atomic layer materials, providing a generic probe of TRSB in two-dimensional superconductors.
In this paper, we set $e=\hbar=k_B=1$ hereafter.

\begin{figure}
\includegraphics[width=75mm]{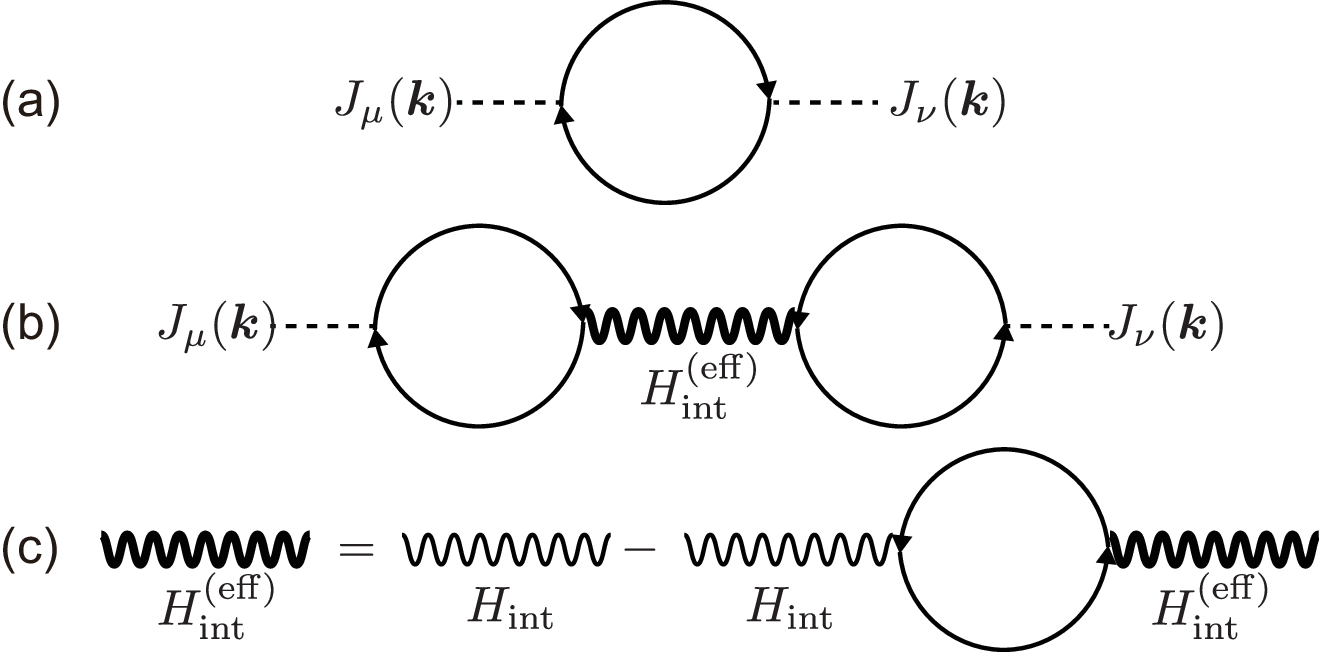}
\caption{Diagrammatic representations of (a,b) the current-current correlation functions and (c) the effective interaction within the random phase approximation (RPA). 
Here, the dotted lines denote the electric current $J_{\mu}(\bm k)$ and $J_{\nu}(\bm k)$, the solid lines are the Gor'kov Green functions, and the thick (thin) wavy lines indicate the effective interaction $H_{\rm int}^{(\rm eff)}$ (the bare interaction $H_{\rm int}$).}
\label{fig:correlation_func}
\end{figure}

{\it Optical sheet conductivity and magneto-optical response.---}
We consider a thin-film chiral superconductor on a substrate of refractive index $n$.
We assume normal incidence and that the film thickness is smaller than the magnetic penetration depth.
In this  limit, the Kerr ($\theta_{\rm K}(\omega)$) and Faraday ($\theta_{\rm F}(\omega)$) rotation angles are given by~\cite{suppl}
\begin{subequations}
\beq
\label{kerrangle}
\theta_{\rm K}(\omega)&=&\frac{1}{2}{\rm arg}\left[\frac{1-n-Z_0\sigma_+(\omega)}{1-n+Z_0\sigma_+(\omega)} \frac{1-n+Z_0\sigma_-(\omega)}{1-n-Z_0\sigma_-(\omega)}\right],\\
\label{faradayangle}
\theta_{\rm F}(\omega)&=&\frac{1}{2}{\rm arg}\left[\frac{1-n+Z_0\sigma_-(\omega)}{1-n+Z_0\sigma_+(\omega)} \right],
\eeq
\end{subequations}
with $\sigma_\pm(\omega)=\sigma_{xx}(\omega)\pm i\sigma_{xy}(\omega)$, where $\sigma_{xx}(\omega)$ and $\sigma_{xy}(\omega)$ denote the longitudinal and zero-field Hall optical sheet conductivities, respectively.
The conductivity tensor is assumed to satisfy $\sigma_{xx}(\omega)=\sigma_{yy}(\omega)$ and $\sigma_{xy}(\omega)=-\sigma_{yx}(\omega)$.
The optical sheet conductivity is defined as the three-dimensional optical conductivity multiplied by the film thickness, and $Z_0=\sqrt{\mu_0/\epsilon_0}$ is the vacuum impedance.
From Eqs.~\eqref{kerrangle} and \eqref{faradayangle}, it follows that both the Kerr and Faraday rotation angles vanish, $\theta_{\rm K}(\omega)=\theta_{\rm F}(\omega)=0$, whenever $\sigma_{xy}(\omega)=0$ as $\sigma_+(\omega)=\sigma_-(\omega)$.

In the superconducting state, the optical conductivity consists of a dissipationless superfluid part $\sigma_{\mu \nu}^{\rm (sf)}(\omega)$ and a regular part $\sigma_{\mu \nu}^{\rm (reg)}(\omega)$~\cite{tinkham_SC},
\beq
\sigma_{\mu \nu}(\omega)=\sigma_{\mu \nu}^{\rm (sf)}(\omega)+\sigma_{\mu \nu}^{\rm (reg)}(\omega),
\eeq
with $\mu,\nu=x,y$.
The former is determined by the superfluid weight $D_{\mu \nu}$ and is given by $\sigma_{\mu \nu}^{\rm (sf)}(\omega)=iD_{\mu \nu}/(\omega+i\eta)$, where $\eta$ is the level-broadening parameter ($\eta\to +0$ in clean systems)~\cite{liang_superfluidweight,hirobe2025anomalous}.
Because ${\rm Re}[\sigma_{\mu \nu}^{\rm (sf)}(\omega)]=\pi D_{\mu \nu}\delta(\omega)$, the superfluid contribution affects magneto-optical responses at microwave frequencies only through the $1/\omega$ tail of ${\rm Im}[\sigma_{\mu \nu}^{\rm (sf)}(\omega)]=\mathcal{P}[D_{\mu \nu}/\omega]$.

We assume the clean limit for the superconductor and evaluate the regular part using the Kubo formula.
The current--current correlation function consists of the bare bubble diagram [Fig.~\ref{fig:correlation_func}(a)], which describes the quasiparticle response, and the two-bubble diagram including the effective interaction, evaluated within the random phase approximation (RPA) [Fig.~\ref{fig:correlation_func}(b,c)].
The poles of the effective interaction give the energies of the collective modes [Eq.~\eqref{effective_interaction}].
Derivations and explicit expressions for $\theta_{\rm K}(\omega)$, $\theta_{\rm F}(\omega)$, and $\sigma_{\mu \nu}(\omega)$ are given in the Supplemental Material~\cite{suppl}.

\begin{figure}[t]
\includegraphics[width=75mm]{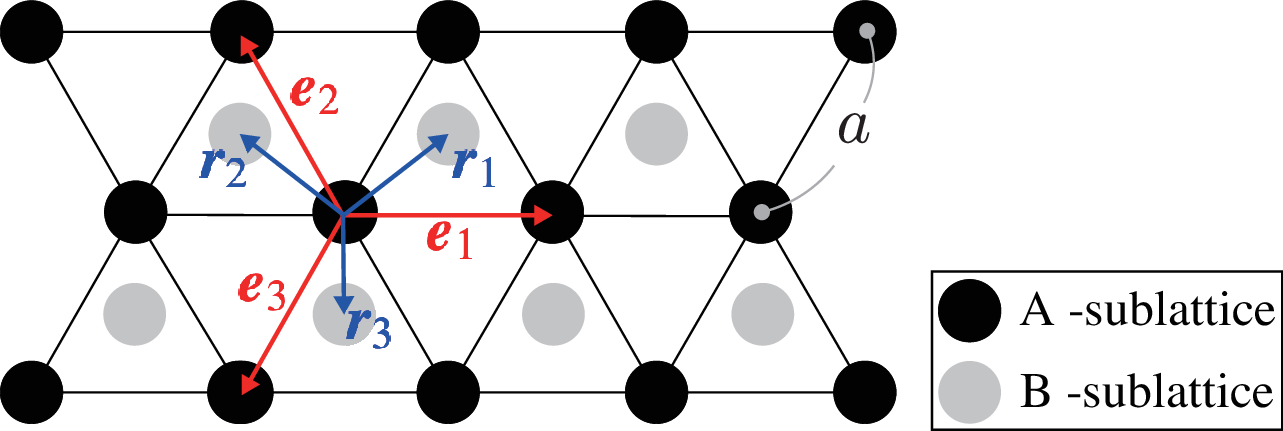}
\caption{Crystal structure of the lattice model described by Eq.~\eqref{HBdG}. The black and grey circles represent the sublattices A and B, respectively.}
\label{fig:lattice}
\end{figure}

{\it Model.---}
As a minimal model of a multiband chiral superconductor belonging to the $E$-irreducible representation, we consider the chiral superconductivity on a two-dimensional triangular lattice with two inequivalent sites (A and B) per unit cell [Fig.~\ref{fig:lattice}].
The Bogoliubov-de Gennes (BdG) Hamiltonian is~\cite{sigristueda}
\beq
\label{HBdG}
\mathcal{H}_{\rm BdG}(\bm k)=\begin{pmatrix}
\mathcal{H}_0(\bm k)&& \Delta(\bm k)\\
\Delta^\dagger (\bm k)&&-\mathcal{H}^{\mathrm{T}}_0(-\bm k)
\end{pmatrix},
\eeq
with $\bm k = (k_x, k_y)$ being the crystal momentum.
Here, $\mathcal{H}_0(\bm k)$ and $\Delta(\bm k)$ are the single-particle Hamiltonian and the pair potential, respectively.

\begin{figure*}
\includegraphics[width=160mm]{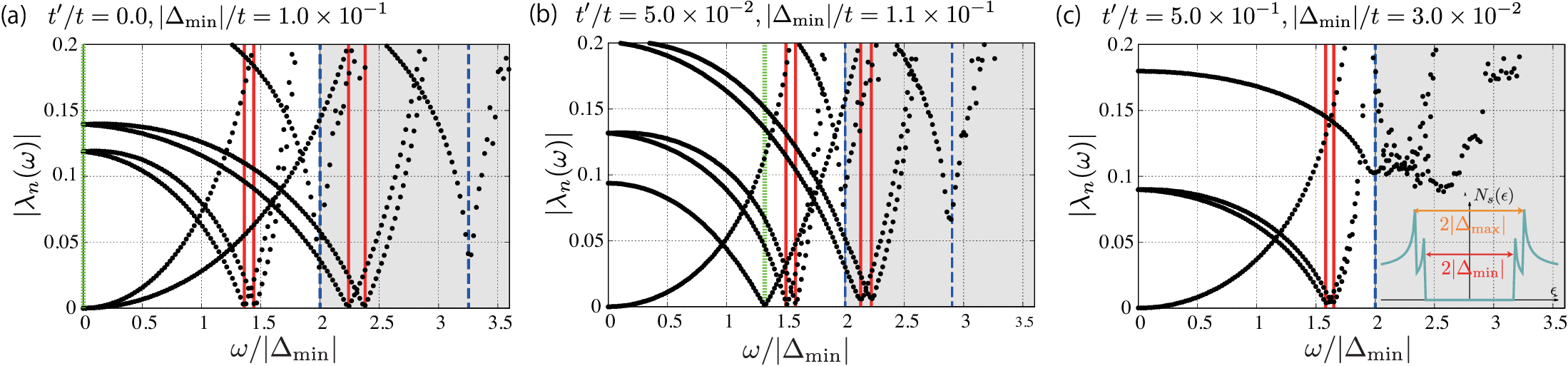}
\caption{The $\omega$ dependence of the absolute values of the eigenvalues $\{\lambda_n (\omega)\}$ (dimensionless) of $1 + |V_{\rm int0}| \Pi^{\rm R}(\omega)$ for $t'/t=0.0, 5.0\times 10^{-2}, 5.0\times 10^{-1}$.
In all panels (a-c), we set $t=1.25$ meV (bandwidth $E_{\mathrm B}\simeq8t=10\;\mathrm{meV}$; equivalently $E_{\mathrm B}/k_{\mathrm B}\simeq116\;\mathrm K$), $a=5$ \AA, $T=0$ K, $M_z=0.5t$, $|V_{\rm int0}|=0.9t$, $\mu=2t$, and $\eta=1.0\times 10^{-4}t$.
The gray shaded area indicates the quasiparticle continuum. 
The vertical lines indicate the Leggett (green, dotted), Higgs (blue, dashed), and clapping (red, solid) modes.
Inset of panel (c) indicates the schematic image for the quasiparticle DOS, $N_s(\epsilon)$, with the smaller (larger) quasiparticle excitation gaps $2|\Delta_{\rm min}|$ ($2|\Delta_{\rm max}|$).}
\label{fig:CM_energy}
\end{figure*}

The normal-state Hamiltonian reads
\beq  
\label{H0}  
\mathcal{H}_0(\bm k) &=& \xi(\bm k) s_0 \sigma_0+{\rm Re}[T(\bm k)] s_0 \sigma_x \nn\\
&-& {\rm Im}[T(\bm k)] s_0 \sigma_y+M_z s_0 \sigma_z.
\eeq  
Here, $s_0$ and $\sigma_0$ are identity matrices, and $\bm s=(s_x,s_y,s_z)$, $\bm \sigma=(\sigma_x,\sigma_y,\sigma_z)$ are the Pauli matrices in spin and sublattice space, respectively.
The first term denotes the intra-sublattice hopping energy $\xi(\bm k) = 2t \sum_{j=1}^3 \cos (\bm k \cdot \bm e_j) - \mu$, measured from the chemical potential $\mu$, where $\bm e_1 = (a, 0)$, $\bm e_2 = \left(-a/2, \sqrt{3}a/2\right)$, and $\bm e_3 = \left(-a/2, -\sqrt{3}a/2\right)$ point to the nearest neighbor atoms [Fig.~\ref{fig:lattice}].
The multiband nature emerges from inter-sublattice hopping term $T(\bm k) = 2t' \sum_{j=1}^3 e^{i \bm k \cdot \bm r_j}$, where $\bm r_1 = \left(a/2, a/(2\sqrt{3})\right)$, $\bm r_2 = \left(-a/2, a/(2\sqrt{3})\right)$, and $\bm r_3 = \left(0, -a/\sqrt{3}\right)$.
In Eq.~\eqref{H0}, $M_z$ is an on-site energy difference between the $\mathrm{A}$ and $\mathrm{B}$ sites, which lifts the band degeneracy, and arises from inequivalent A and B atoms or from substrate-induced sublattice asymmetry.

The spin-triplet chiral pair potential is given by
\beq  
\label{eqgap}  
\Delta(\bm k) &=& \left(p_1(\bm k) + i p_2(\bm k)\right) \left( \Delta_{\rm A} s_x  \sigma_{\rm A} + \Delta_{\rm B} s_x  \sigma_{\rm B} \right),  
\eeq  
where $\sigma_{\rm A} = (\sigma_0 + \sigma_z)/2$ and $\sigma_{\rm B} = (\sigma_0 - \sigma_z)/2$ are the projection operators, and $\Delta_{\rm A}$ and $\Delta_{\rm B}$ denote the gap amplitudes for the A and B sites.
$p_1(\bm k)=c_1\left(\sin k_xa + \sin \frac{k_xa}{2} \cos \frac{\sqrt{3}}{2} k_ya\right)$ and $p_2(\bm k) =c_2\cos \frac{k_xa}{2} \sin \frac{\sqrt{3}}{2} k_ya$ are the two basis functions of the $E$ irreducible representation, where $c_1\simeq 1.07$ and $c_2\simeq 1.86$ are normalization constants chosen to satisfy $(1/N)\sum_{\bm k} p_1^2(\bm k)=(1/N)\sum_{\bm k} p_2^2(\bm k)=1$~\cite{sigristueda}.
With this normalization, $p_1(\bm{k})+ip_2(\bm{k})\propto k_x+ik_y$ near  $\bm{k}=0$.
Accordingly, Eq.~\eqref{eqgap} describes a chiral $p$-wave order parameter, consistent with the three-fold rotational symmetry of the lattice.

To describe the chiral state in the mean-field theory, we introduce a phenomenological attractive interaction
~\cite{sigristueda,suppl}
\beq
\label{Hint_ph}
H_{\rm int}
&=&-\frac{1}{4N}\sum_{\bm k,\bm k'}\sum_{j=1}^2\sum_{\alpha=x,y}|V_{\rm int0}|
\left[\psi^\dag(\bm k)
\Gamma_{j\alpha}(\bm k)
\psi (\bm k)\right]\nn\\
&\times&\left[
\psi^\dag(\bm k')
\Gamma_{j\alpha}(\bm k') 
\psi(\bm k')\right],
\eeq
with $\Gamma_{1\alpha}(\bm k)=p_1(\bm k)s_x \sigma_0 \tau_\alpha$ and $\Gamma_{2\alpha}(\bm k)=p_2(\bm k)s_x \sigma_0 \tau_\alpha$.
$\psi(\bm k)$ is the Nambu spinor and $\bm \tau=(\tau_x,\tau_y,\tau_z)$ are the Pauli matrices in the particle-hole space.
To obtain the chiral ground state, we fix the relative phase between the two basis functions to $\pi/2$ and solve the gap equation self-consistently.
Although the pairing interaction is sublattice-independent, the self-consistent gaps $\Delta_{\rm A,B}$ become inequivalent, $\Delta_{\rm A}\neq \Delta_{\rm B}$, when $M_z\neq 0$ due to sublattice asymmetry.
This minimal model is directly related to candidate chiral superconductors, such as homobilayer/twisted-bilayer TMDCs~\cite{yuantmd,kanasugitmd,wutmd}.

{\it Collective excitations.---}
The two-bubble diagram with the effective interaction [Fig.~\ref{fig:correlation_func} (b,c)] must be taken into account for the self-consistent treatment of the pair potential and its fluctuations due to incident light~\cite{Nambu_gaugeinvariance,watanabe_opticalresponse,watanabe2025gauge,tanaka2025vertex,kamatanileggett,nagashimaleggett}.
Within the RPA, the effective interaction is given by~\cite{suppl}
\begin{subequations}
\label{effective_interaction}  
\begin{align}
H^{(\rm eff)}_{\rm int}(i\omega_m)&=-\frac{1}{4N}\sum_{\bm k,\bm k'}\sum_{j,l=1}^2\sum_{\alpha,\beta=x,y}
V^{(j\alpha)(l\beta)}_{\rm eff}(i\omega_m)
\nn\\
&\times \left[\psi^\dag(\bm k)
\Gamma_{j\alpha}(\bm k)
\psi (\bm k)\right]
\left[
\psi^\dag(\bm k')
\Gamma_{l\beta}(\bm k') 
\psi(\bm k')\right],\\
V_{\rm eff}(i\omega_m)&=|V_{\rm int0}|\left[1 + |V_{\rm int0}| \Pi^{\rm M}(i\omega_m)\right]^{-1},
\end{align}
\end{subequations}
where $V_{\rm eff}(i\omega_m)$ is a $4\times4$ matrix in the channel space spanned by $j,l=1,2$ (two basis functions $p_{1,2}(\bm k)$) and $\alpha,\beta=x,y$ (two particle-hole channels $\tau_{x,y}$). 
$\Pi^{\rm M}(i\omega_m)$ is a matrix in the channel space, and its matrix elements are given by 
\beq  
\left[\Pi^{\rm M}(i\omega_m)\right]^{(j\alpha)(l\beta)} &=& \int \frac{d^2k}{(2\pi)^2}\sum_{n_1,n_2}\frac{f_{n_1n_2}\Gamma^{n_1n_2}_{j\alpha}\Gamma^{n_2n_1}_{l\beta}}{i\omega_m+\epsilon_{n_1}-\epsilon_{n_2}},  
\eeq  
with $f_{n_1n_2}(\bm k)=f(\epsilon_{n_1}(\bm k))-f(\epsilon_{n_2}(\bm k))$, where $f(\epsilon)=(1+e^{\epsilon/T})^{-1}$ is the Fermi-Dirac distribution function and $\epsilon_n(\bm k)$ is the quasiparticle energy defined by $\mathcal{H}_{\rm BdG}(\bm k)|\varphi_n(\bm k)\rangle =\epsilon_n(\bm k)|\varphi_n(\bm k)\rangle$.
Here, $\Gamma^{n_1n_2}_{j\alpha}=\langle \varphi_{n_1}(\bm k)|\Gamma_{j\alpha}(\bm k)|\varphi_{n_2}(\bm k)\rangle$ denotes the quasiparticle-representation of the basis functions.

For all parameter sets except $t'=0$ and $M_z=0$, the normal-state band structure hosts two distinct Fermi surfaces surrounding the $\Gamma$ point. In the weak-coupling regime considered here, projecting the order parameter defined in the sublattice basis onto the band basis yields an almost isotropic gap magnitude on each Fermi surface, while retaining the chiral phase winding.

The energies ($E_{\rm cmd}$) and damping rates ($\gamma_{\rm cmd}$) of the collective modes correspond to zeros of the denominator of the effective interaction, $1 + |V_{\rm int0}| \Pi(z)$, in the complex $z$ plane, $z=E_{\rm cmd}+i\gamma_{\rm cmd}$, where $\Pi(z)\equiv\Pi^{\rm M}(i\omega_m\to z)$.
We identify them from sharp minima of the eigenvalues, $|\lambda_n(\omega)|$, of the $4\times 4$ matrix denominator on the real-frequency axis, $1 + |V_{\rm int0}| \Pi^{\rm R}(\omega)$ with $\Pi^{\rm R}(\omega)=\Pi^{\rm M}(i\omega_m\to \omega+i\eta)$.
Such minima indicate long-lived modes ($\gamma_{\rm cmd}\ll E_{\rm cmd}$).

Figure~\ref{fig:CM_energy} (a--c) shows $|\lambda_n(\omega)|$ as a function of $\omega$ for $t'/t=0.0, 5.0\times 10^{-2}, 5.0\times 10^{-1}$.
As shown in the inset of Fig.~\ref{fig:CM_energy} (c), the quasiparticle density of states (DOS) exhibits two coherence peaks due to the inequivalence of the two sublattices.
The frequencies in all panels are scaled by the smaller energy gap $|\Delta_{\rm min}|$, and the gray shaded area indicates the quasiparticle continuum.
In all panels, the Nambu-Goldstone (NG) mode (global phase oscillation) appears at zero frequency (i.e., $|\lambda_n(0)|=0$).
Its energy is pushed up to the plasma frequency (far above $2|\Delta_{\rm min}|$) via the Anderson-Higgs mechanism once the long-range Coulomb interaction is included~\cite{andersongauge,andersonRPA,higgsgauge}.

With $p_\pm (\bm k)=p_x (\bm k)\pm ip_y (\bm k)$, the general fluctuation of the order parameters can be written as
\beq
\delta \left(\Delta_{\rm A,B} p_+(\bm k) \right)= \delta \Delta_{\rm A,B}^{(1)}p_+(\bm k)+\delta \Delta_{\rm A,B}^{(2)}p_-(\bm k)\,.
\label{eq:OP_fl}
\eeq
Here, $\delta \Delta_{\rm A,B}^{(1)}$ preserves the ground-state chirality $m_z$ and yields NG, Higgs (amplitude), and Leggett modes, while $\delta \Delta_{\rm A,B}^{(2)}$ mixes in the opposite chirality and defines the clapping modes~\cite{vollhardt2013superfluid}. 
In Fig.~\ref{fig:CM_energy} (a--c), the vertical lines indicate the Leggett (green), Higgs (blue), and clapping (red) modes, respectively.
The Leggett mode -- the relative phase oscillation between $\Delta_{\rm A}$ and $\Delta_{\rm B}$ -- becomes massless when the intersublattice hopping is switched off, $t'=0$~\cite{leggett1966number,murotanileggett} [Fig.~\ref{fig:CM_energy} (a)], but acquires a mass for $t'\neq 0$ [Fig.~\ref{fig:CM_energy} (b)], and is pushed above $2|\Delta_{\rm min}|$ once $|t'|$ exceeds a threshold value [Fig.~\ref{fig:CM_energy} (c)].
The energies of the Higgs modes coincide with the two coherence peaks of the quasiparticle DOS, and acquire a finite lifetime ($|\lambda_n(\omega)|\neq 0$) due to the decay of Cooper pairs into electrons and holes.

\begin{figure}[t]
\includegraphics[width=70mm]{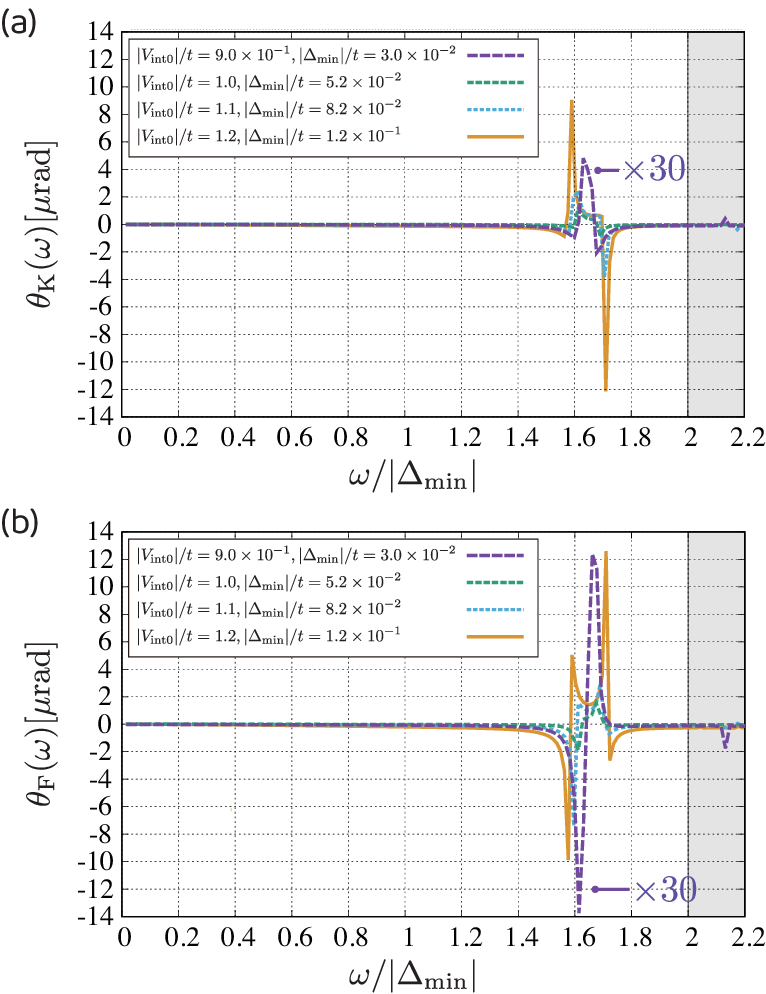}
\caption{The $\omega$-dependence of (a) the Kerr rotation angle and (b) Faraday rotation angle for $|V_{\rm int0}|/t=9.0\times 10^{-1},1.0,1.1,1.2$.
The gray shaded area indicates the quasiparticle continuum. 
In all panels, we set $t=1.25\;\mathrm{meV}$ (bandwidth $E_{\mathrm B}\simeq8t=10\;\mathrm{meV}$), $a=5\;\text{\AA}$, $T=0\;\mathrm K$, $t'=M_z=0.5t$, $\mu=2t$, $\eta=1.0\times10^{-4}\;t$, and $n=1.45$.}
\label{fig:kerrangle}
\end{figure}

The clapping modes in Eq.~\eqref{eq:OP_fl} are the relative phase and amplitude oscillations between the two chiral components.
For two sublattices, therefore, there are four distinct clapping modes [Fig.~\ref{fig:CM_energy} (a,b)].
In two dimensions, chiral superconductors are fully gapped. 
Figure~\ref{fig:CM_energy} (a--c) shows that two of them lie well below the pair-breaking threshold $2|\Delta_{\rm min}|$, whereas the other two lie above $2|\Delta_{\rm min}|$. 
These results show that the clapping modes remain below the quasiparticle excitation gap over a wide range of parameters, as long as the two chiral components $(k_x \pm i k_y)^{m_z}$ are degenerate within a two-dimensional irreducible representation, even in multiband systems.

In single-band chiral superconductors, the energies of the clapping modes are determined by the gap amplitudes $|\Delta|$ and by the Fermi surface anisotropy~\cite{sauls_clapping}.
For an isotropic Fermi surface, the phase-like and amplitude-like clapping modes are degenerate at $\omega=\sqrt{2}|\Delta|$.
In our model, even when the sublattices are decoupled ($t'=0$), the energies of the clapping modes deviate from $\sqrt{2}|\Delta_{\rm min}|$ and $\sqrt{2}|\Delta_{\rm max}|$ due to the Fermi surface anisotropy.

As shown in Fig.~\ref{fig:CM_energy} (a) for $t'=0$, the two clapping modes with $\omega>2|\Delta_{\rm min}|$ have infinite lifetimes because they decouple from quasiparticles in the window $|\Delta_{\rm min}|<|\epsilon|<|\Delta_{\rm max}|$.
Once the intersublattice hopping is switched on, the high-energy clapping modes become strongly damped. 
At the same time, the intersublattice hopping introduces a finite lifetime for the lower-energy clapping modes ($\omega<2|\Delta_{\rm min}|$), yet they remain long-lived ($|\lambda_n(\omega)|\ll1$), indicating their stability in multiband systems [Fig.~\ref{fig:CM_energy} (c)].

Because the superconducting energy scale is much smaller than any other normal-state energy scale, we focus on the regime $|\Delta_{\rm A,B}|\ll |t'|$, where the energy of the Leggett modes lies far above the quasiparticle continuum threshold $2|\Delta_{\rm min}|$.

{\it Polar Kerr and Faraday effects.---}
With $V_{\rm eff}(i\omega_m)$ obtained above within the RPA, we evaluate the bare bubble [Fig.~\ref{fig:correlation_func}(a)] and two-bubble [Fig.~\ref{fig:correlation_func}(b)] contributions to the optical conductivity tensor, and compute the Kerr and Faraday rotation angles.
Figure~\ref{fig:kerrangle}(a,b) shows the resonant enhancements and the sign reversals of the Kerr and Faraday rotation angles as functions of the light frequency (anti-Lorentzian line shape).
The resonances occur when the light frequency matches the energies of the lower-energy clapping modes, thereby probing them directly. 
The calculated results for the longitudinal and Hall optical sheet conductivities are provided in the Supplemental Materials~\cite{suppl}.

Note that, in clean single-band chiral superconductors, the collective-mode contribution to the zero-field optical Hall conductivity vanishes.
Hence, the collective-mode-induced PKE and FE are unique to multiband chiral superconductors.
A spatially uniform electric field accelerates only the center-of-mass motion of the condensate, but does not couple to its relative orbital motion (polarized orbital angular momentum) that underlies broken time-reversal and mirror-reflection symmetries~\cite{readFQHE}.
In clean single-band chiral superconductors with Galilean invariance, these two motions are decoupled, resulting in $\sigma_{xy}(\omega)=0$ and thus $\theta_{\rm K}(\omega)=\theta_{\rm F}(\omega)=0$~\cite{com1}.
By contrast, in multiband chiral superconductors, interband hybridization (e.g., intersublattice hopping) breaks this exact separation and allows the PKE and the FE.

The quasiparticle contribution to the Kerr and Faraday rotation angles, which would onset at $\omega=2|\Delta_{\rm min}|$, vanishes in our model.
In the clean limit, the quasiparticle Hall response requires interband hybridization and the presence of the time-reversal-odd bilinears (TROBs)~\cite{Denys_TROB, Brydon_TROB}.
As shown in the Supplemental Materials, the TROBs vanish for this model, indicating the absence of the quasiparticle contribution to $\sigma_{xy}(\omega)$~\cite{suppl}.
Hence, in our model, the PKE and the FE arise solely from the collective modes.

Although the quasiparticle Hall response ($\sigma_{xy}^{\rm (QP)}(\omega)$) is finite in realistic materials, the contribution of the low-energy clapping modes  ($\sigma_{xy}^{\rm (CM)}(\omega)$) can be isolated because it lies within the quasiparticle excitation gap ($\omega<2|\Delta_{\rm min}|$).
In general, the relative magnitudes are estimated as~\cite{suppl}
\beq
\frac{{\rm max}_{\omega}|\sigma_{xy}^{\rm (CM)}(\omega)|}{{\rm max}_{\omega}|\sigma_{xy}^{\rm (QP)}(\omega)|}=\mathcal{O}\left(
\frac{V_{\rm pair}}{\Delta}\right),
\eeq
where $V_{\rm pair}$ is the bare pairing interaction and $\Delta$ is the superconducting gap.
This ratio is typically $10$--$100$ in candidate chiral superconductors.

The peak values of the Kerr and Faraday rotation angles depend on $|\Delta_{\rm min}|/t$, while the resonant peak widths are given by $\eta$.
The peak decreases as $|\Delta_{\rm min}|/t$ becomes smaller; nevertheless, even for $|\Delta_{\rm min}|/t=3.0\times 10^{-2}\text{--}1.2\times 10^{-1}$ (with band width $E_{\rm B}\simeq 8t$), they remain the order of 100 nrad--10 $\mu$rad, well within the experimental sensitivity~\cite{xia_PKE, shemm_PKE_uru2si2,schemm2014observation,hayes_kerr}.
Such values of $|\Delta_{\rm min}|/t$ are achievable in uranium-based superconductors, and even larger ratios are expected in superconducting atomic layer materials.
Therefore, observation of the microwave polar Kerr and Faraday effects is feasible in these systems.

We emphasize that microwave Kerr/Faraday resonances not only detect TRSB but also spectrally resolve the characteristic collective excitations of the chiral superconducting condensate.
Therefore, it is generally advantageous to measure these magneto-optical responses at microwave frequencies rather than in the near-infrared frequencies.

{\it Conclusion.---}
We investigated the polar Kerr and Faraday effects in two-dimensional multiband chiral superconductors.
We showed the stability (long lifetime) of the clapping modes, a collective mode characteristic of chiral superconductors, and demonstrated that they give rise to magneto-optical responses with resonant enhancement in the microwave regime.
These effects are directly accessible in superconducting atomic layer materials--such as homobilayer/twisted-bilayer TMDCs and multilayer graphene--thereby providing generic probes of chiral superconductivity in two dimensions~\cite{yuantmd,kanasugitmd,wutmd,geier2024chiral}.

{\it Note added.---}
The collective mode contribution to the zero-field optical Hall conductivity was reported in Ref.~\cite{niederhoff2025current}; however, the polar Kerr and Faraday effects were not evaluated in this work.

{\it Acknowledgements.---}
TM appreciates N. Tsuji, T. Mizushima, S. Fujimoto, S. Watanabe, A. Daido, K. Shinada, H. Tanaka, Y. Hirobe, and C.-g. Oh, for helpful discussions.
TM thanks K. Hashimoto and K. Ishihara for motivating this work from an experimental point of view.
TM was supported by JSPS KAKENHI (Grant No. JP24KJ0130 and JP24H00007) and JST CREST (Grant No. JPMJCR19T2). 
YY was supported by JSPS KAKENHI (Grant Numbers JP22H01181, JP22H04933, JP23K22452, JP23K17353, JP24K21530, JP24H00007, JP25H01249).

\bibliographystyle{apsrev4-1_PRX_style}
\bibliography{chiral.bib}
\end{document}